\documentclass{article}

\usepackage{amsmath}
\usepackage{amssymb}
\usepackage{euscript}
\usepackage{sprocl}

\def\lint{\int\limits}
\def\bbeta{\bar{\beta}}

\newcommand{\Rcal}{{\cal R}}

\def\be{\begin{equation}}
\def\ee{\end{equation}}
\def\bea{\begin{eqnarray}}
\def\eea{\end{eqnarray}}

\begin{document}

\rightline{\bf CERN-TH/99-06}
\rightline{\bf UTHEP-98-1001}
\rightline{\bf January 1999}

\vspace{4mm}

\title{PRECISION CALCULATIONS OF HEAVY BOSON PRODUCTION - YFS MONTE
CARLO APPROACH}

%
%
%
%
%
%
%
%
%

\author{M. SKRZYPEK$^{ab}$,
   S. JADACH$^{ab}$,
   W. P\L{}ACZEK$^{ea}$,
   B.F.L. WARD$^{acd}$,
   Z. W\c{A}S$^{ba}$ 
}

\address{
$^a$CERN, Theory Division, CH-1211 Geneva 23, Switzerland,\\
 $^b$Institute of Nuclear Physics,
  ul. Kawiory 26a, 30-055 Cracow, Poland,\\
$^c$Department of Physics and Astronomy,\\
  The University of Tennessee, Knoxville, TN 37996-1200,\\
 $^d$SLAC, Stanford University, Stanford, CA 94309,\\
$^e$  Institute of Computer Science, Jagellonian University,\\
        ul. Nawojki 11, 30-072 Cracow, Poland
}
\address{Talk given by M. Skrzypek at IVth International 
Symposium \\ on Radiative Corrections,
Barcelona (Catalonia, Spain), 
Sept.\ 8-12, 1998.}

\maketitle\abstracts{ In this talk I present the current status of the
family of Monte Carlo programs for four-fermion physics based on the
Yennie-Frautschie-Suura resummation of soft real and virtual photons. 
I focus on their applications to LEP2.}

\section{Introduction}

The role of Monte Carlo (MC) codes at LEP2 is more important than at
LEP1. This is primarily because the physics is more complicated: we deal with
a multitude of four-fermion final states, their topology is complicated
and not that many semi-analytical results exist. As a consequence, for
example, the W-mass fits are performed at LEP2 with the help of MC. 
At the time of the LEP2 Workshop in 1995 a list of wishes concerning
the features of the ``ultimate Monte Carlo'' was constructed, and the goal 
of 0.5\% precision tag for the total cross-section for WW-physics was set
\cite{yr:lep2}. As the experiment is already well advanced it is
time to confront these expectations with reality. 
I will do it for the family of four-fermion MC codes
based on the YFS technique: KoralW, YFSWW and YFSZZ developed by our
group. Let me begin with a small technical introduction.

\section{YFS Monte Carlo Approach to Four-Fermion Calculations}

In their classical paper \cite{yfs:1961}, Yennie, Frautschie and Suura (YFS)
reorganized the perturbative series of QED photonic corections for
an arbitrary process in a manifestly infrared (IR) finite way. 
The key step was in the extraction of the universal
real ($\tilde S$) and virtual ($S$) functions containing IR
singularities. After properly integrating and resumming them to {\em all
orders} the remaining new perturbative series in $\bar\beta_n$ functions 
became IR-finite. As compared to the fixed-order approach 
this layout is especially convenient for MC
algorithms. One is not faced
here with problems of IR cut-off, negative
distributions or large real--virtual cancellations. Moreover, one can 
generate a {\em multiple} photonic bremsstrahlung instead of just one or
two photons at most. 
This approach has been pioneered by two of us (SJ and BFLW) in refs.\
\cite{jadach:1988,jadach:1989,yfs2:1990}. 
In mathematical terms, the master formula of YFS
series, adapted to the MC needs, is the following 
\cite{jadach:1988,jadach:1989,yfs2:1990}:
\begin{equation}
\nonumber
\label{master}
\begin{split}
 & \sigma =
  \sum_{n=0}^\infty \frac{1}{n!}
  \lint \prod_{i=1}^4 \frac{d^3q_i}{q_i^0} 
  \left(\prod_{i=1}^n       \frac{d^3k_i}{k^0_i}
    \tilde{S}(p_1,p_2,k_i) \right)
  \delta^{(4)}\bigg(p_1+p_2-\sum_{i=1}^4 q_i
  -\sum_{i=1}^n k_i \bigg)
\\
  &
  \exp\bigg(  2\alpha \Re B
              + \lint \frac{d^3k}{k^0} \tilde{S}(p_1,p_2,k)
              (1-\theta_\epsilon)
      \bigg)
  \Bigg[
         \bbeta^{(3)}_0(p^\Rcal_r,q^\Rcal_s)
        +\sum_{i=1}^n
        \frac{\bbeta^{(3)}_{1}( p^\Rcal_r, q^\Rcal_s, k_i)}{\tilde{S}(k_i)}
\\&
        +\sum_{ i>j }^n
        \frac{\bbeta^{(3)}_{2}( p^\Rcal_r, q^\Rcal_s, k_i, k_j)
                             }{\tilde{S}(k_i)\tilde{S}(k_j)}
        +\sum_{ i>j>l }^n
        \frac{\bbeta^{(3)}_{3}( p^\Rcal_r, q^\Rcal_s, k_i, k_j, k_l)
                             }{\tilde{S}(k_i) \tilde{S}(k_j) \tilde{S}(k_l) }
  \Bigg]
  \Theta^{cm}_\epsilon.
\end{split}
\end{equation}
To explain it very briefly, the $\exp(\dots)$ function 
is the place where IR
virtual ($\Re B$) and real ($\int \tilde S$) singularities cancel;
$\prod d^3k_i \tilde S$ describe the multiple photonic emission
and $\sum \bar \beta_i$ is the perturbative expansion of a non-IR part of 
the matrix element (truncated to the third order). 
To lowest order, $\bar \beta_0$ is just the Born matrix element. 
This formula is the basis of the MC codes that we developed for
four-fermion physics.

\section{The ``Four-Fermion MC Toolbox''}

For the moment we provide three complementary MC codes: 
KoralW, YFSWW and YFSZZ.  I will refer to them together as
``Four-Fermion MC Toolbox''.

 {$\bullet$ \bf KoralW}

The main features of KoralW \cite{koralw:1995a,koralw:1995b,koralw:1998} are: 
(1) it generates all four-fermion final
states with the complete massive Born-level matrix elements and two
pre-samplers for complete, massive, four-fermion phase space. 
The matrix element
comes from the automated package GRACE v.\ 2.0.
(2) Anomalous WWV couplings are included in CC03 graphs.
(3) Multiple initial-state photons with finite $p_T$ are generated by the YFS
technique and a QED ${\cal O}(\alpha^3)$ leading-log 
matrix element is included.
(4) Coulomb effect,`` Naive QCD'' correction and non-diagonal CKM matrix are
included.
(5) JETSET, PHOTOS and TAUOLA are interfaced.
(6) Semi-analytical routine KorWan for CC03 graphs is included.
(7) Analysis of Bose-Einstein effect is done as a stand-alone
application (in C++) \cite{JadachZalewski:1997}.

As compared with the previous version (1.33) the main novelties of v.\ 1.41
are the following:
(1) we have added a second, independent presampler for the four-fermion
phase space, which becomes a very powerful test of the code;
(2) we have added the third-order leading-log correction to the QED ISR matrix
element;
(3) the anomalous couplings can now be parametrized in three different
ways;
(4) any combination of final states can now be specified by the user;
(5) CKM matrix and colour reconnection probability are now input
parameters.

The most difficult part of the code is the phase-space generator.
The origin of the difficulty is in the multitude and complexity of the
available final states. Naively counting one has to deal with $9\times
9$ decay channels of WW (CC processes) and $11\times11$ of ZZ
(NC processes), 202 in total. Each channel can contribute up to 
100 Feynman graphs, so
the total number of ``objects'' to integrate exceeds 10 000.
The solution lies in a {\em multi-branch MC algorithm}. This
approach can already be found in TAUOLA \cite{tauola:1992,tauola:1993} 
and FERMISV \cite{kleiss:1993} MC codes. 
The general formula is the following:
\vspace{ -0.1cm}
\begin{eqnarray}
\sigma &=& \int d{\rm Phsp}\;   \vert M \vert^2 
       = \left< \frac{{ \vert M \vert^2}}{\tilde f_{CR}} \right>_{d\tilde\rho}
           \int d\tilde\rho,\;\;\; 
\nonumber
\\
d\tilde\rho &=&  d{\rm Phsp}\;   \tilde f_{CR}
 = \sum_i^{Branch}   ds_1^ids_2^i \prod_{j=1}^3 d\cos\theta_j^i\;
       d\phi_j^i\; \lambda_j^i \; p_i\; \tilde f_{CR}^i .
\label{DPS}
\nonumber
\end{eqnarray}
\noindent
Skipping details (see \cite{koralw:1998}), let me only say that the
process specific information is hidden in the $\tilde f_{CR}^i$ functions.
These functions must contain all
mass and angular singularities of the Feynman diagrams: $1/s, 
1/[(s-M^2)^2 +M^2\Gamma^2], 1/t, 1/u$ and so on.
All the above simple functions
are used in $\tilde f_{CR}^i$ in the form of subsequent sub-branches
with their own
splitting probabilities. It is a non-trivial task to
fine-tune all these coefficients to get the efficient modelling of all 
the matrix elements.

{$\bullet$ \bf YFSWW}

Starting from the same master formula, the YFSWW code
\cite{yfsww2:1996,yfsww3:1998,yfsww3:1998b} features
the ${\cal O}(\alpha)$ corrections to W-pair production:
(1) CC03 (signal) graphs are included in the matrix element;
(2) first-order corrrections for WW production process are included
in the exact form of refs.\ 
\cite{fleischer:1989,kolodziej:1991%
} 
as well as in an improved Born
approximation of ref.\ \cite{dittmaier:1992};
(3) YFS-type bremsstrahlung is generated from both initial and
intermediate (WW) states;
(4) anomalous couplings are included in the Born matrix element;
(5) JETSET, PHOTOS and TAUOLA are interfaced.

On the technical side, YFSWW presents a number of new ideas. First
of all the YFS scheme originally derived for fermions had to be
extended to bosons. In a nutshell this is possible since soft photons 
are ``blind'' to
spin. Both the fermionic and bosonic vertices look the
same in the IR limit:
\begin{equation}
\begin{split}
\lim_{k\to 0}&\left[
 \cdots \frac{i(-iQ_ee\gamma^\mu)}{\not\!p'-m_e+i\epsilon}u(p)\right]
=\cdots \frac{(Q_ee)(2p^\mu-k^\mu)}{k^2-2kp+i\epsilon}u(p),
\;\;\;\;\;\;\;\;\;\;\;\;\;\;\;\;\;\; \hbox{[fermions]}
\\
%
\lim_{k\to 0}&
\biggl[
 \cdots \frac{(-i)(g^{\alpha''\alpha'}-{p'}^{\alpha''}{p'}^{\alpha'}/M_W^2)
                    }{{p'}^2-M_W^2+i\epsilon}(iQ_We)
               [g_{\alpha'\beta}(2p-k)^\mu+g_{\alpha'}^\mu(-p+2k)_\beta
\notag \\ &
               +g^\mu_\beta(-p'-2k)_{\alpha'}]\epsilon_-^\beta(p) 
       \biggr]
=\cdots \frac{(Q_We)(2p^\mu-k^\mu)
           }{k^2-2kp+i\epsilon}\epsilon_-^{\alpha''}(p).
\;\;\;\;\; \hbox{[vect.\ bosons]}
\label{eq5}
\end{split}  
\end{equation}
The second important extension of the original YFS paper \cite{yfs:1961} 
was in deriving
the YFS form factors in the case of heavy massive particles, contrary to the
original small mass approximation of ref.\ \cite{yfs:1961}. 
These functions can
be found in \cite{yfsww2:1996}.
Next, since photons are radiated from the W-bosons with a finite width,
one must do it in a way that respects gauge invariance. We did it by
adding compensating loop corrections that restore gauge invariance in a
similar way as in refs \cite{baur:1995,argyres:1995}.
Finally, we had to avoid double counting of Coulomb effect in the matrix
element and in YFS virtual B-function, both arising from the same type
of loop corrections. We have done that by a proper
redefinition of the B-function. 

With the help of the YFSWW code we can evaluate the size of 
${\cal O}(\alpha)$ corrections, see the table below 
(on the example of the $c\bar s e\bar\nu_e$ final state).
\begin{table}[!ht]
\centering
\begin{tabular}{||c|c|c|c||}
\hline\hline
 $E_{CM}$ [GeV]  & $\sigma_0$ [pb]  
& $(\sigma_1^{ex} - \sigma_1^{LL})/\sigma_0$
& $(\sigma_1^{ex} - \sigma_1^{ap})/\sigma_0$ 
\\
\hline\hline
$161$ & $0.1768$ & $-0.83\%$ & $+0.22\%$ \\
$175$ & $0.5891$ & $-1.32\%$ & $-0.006\%$ \\
$190$ & $0.6792$ & $-1.71\%$ & $-0.22\%$ \\
$205$ & $0.6850$ & $-2.22\%$ & $-0.61\%$ \\
\hline
$500$ & $0.2710$ & $-4.72\%$ & $-3.09\%$ \\
\hline\hline
\end{tabular}
\label{tab:approx}
\end{table}
\vskip -0.1truecm
The main novelty of this result is that the corrections are calculated
within the realistic full four-fermion MC framework, 
allowing for any experimental cuts.
The numbers show that the ${\cal O}(\alpha)$ correction, of
about 2\%, is
similar in size to the $e^+e^- \to W^+W^-$ case analysed in the
literature, see \cite{yr:lep2} and refs.\ therein.
Therefore it must be included in the MC in order to reach the
targeted 0.5\% precision level. The approximate Born cross-section
barely fits within the 0.5\% limit, depending on the actual highest
energy available at LEP2.

The last question to be addressed here is how to combine various
physical corrections of KoralW and
YFSWW? This almost trivial question requires some attention
\cite{koralw:1997} since certain effects, if they are put together improperly, 
can lead to severe effects, such as the case of running W-width 
in background graphs \cite{argyres:1995}.
The natural way out is to use a ``Multiparameter Linear Interpolation'',
that is just a series expansion around the ``minimal'' CC03 cross-section:
$
\sigma_{MLI}=\sigma_{CC03} +(\sigma_{CCall}- \sigma_{CC03})
 +(\sigma_{{\cal O}(\alpha)}- \sigma_{CC03}) 
+ \dots (\Gamma(s), \hbox{ACC, NQCD},\dots).
$
As the above expression has quite a few terms it
becomes impractical in actual usage. We have shown
in ref.\ \cite{koralw:1997} an example of the total cross section (with
some cuts) that one can equivalently use a 
``Common Sense Interpolation'' by switching on all available corrections
in the same MC run, provided one does some minimal cross-checks for
each type of observables. This way the above expresion would be reduced
to two corrections only (from KoralW and YFSWW), and one day, if also
${\cal O}(\alpha)$ corrections are included in KoralW, 
to just one MC run.

{$\bullet$ \bf YFSZZ}

YFSZZ \cite{yfszz:1997} is a dedicated code for the production 
and decay of Z-pairs in the
process $e^+e^- \to ZZ \to f_1\bar f_1 f_2 \bar f_2$. It features:
(1) signal NC02 (ZZ doubly resonant) matrix element; 
(2) anomalous ZZV couplings;
(3) multiple YFS-based bremsstrahlung with an ${\cal O}(\alpha^2)$
leading-log QED matrix element.
This code is yet to be explored and developed as the physics
of Z-pairs progresses.

\vskip -0.2truecm
\section{Precision of ``Four-Fermion Toolbox''}

We can now turn to the basic question: What is the overall precision of
the (KoralW+YFSWW) ``Toolbox'' for the WW physics at LEP2? 
As stated earlier, we focus
on the total cross-section with an eye on a 0.5\% precision level.

{$\bullet$ \bf KoralW}

(1) {\it Technical precision} 
    is estimated at 0.2\% based on: 
    (a) internal comparisons of two presamplers, 
    (b) comparisons with other codes \cite{yr:lep2,koralw:1997},
    (c) comparisons with semi-analytical results of the KorWan code
    (for CC03 matrix element).

(2) {\it Physical precision} is estimated at 2\% based on the size
    of the ${\cal O}(\alpha)$ correction calculated by the YFSWW code.

{$\bullet$ \bf YFSWW}

(1) {\it Technical precision}
    is estimated at 0.2\% by comparing results of two technically different
    implementations of the code: YFSWW-2 and YFSWW-3 as well as
    comparison with KoralW.

(2) {\it Physical precision} is still under study. Preliminary
    estimate of 0.5\% uses in particular 
    the fact that the
    so-called ``non-factorizable'' corrections 
    are negligible. This was first
    pointed out by Fadin, Khoze and Martin 
    \cite{fadin:1993,fadin:1994,fadin:1994b} and then worked
    out in detail by other groups 
    \cite{melnikov:1996,beenakker:1997,beenakker:1997b,%
dittmaier:1998,dittmaier:1998b}.

{$\bullet$ \bf Total}

Collecting the above uncertainties, we can give the {\em
preliminary} estimate of the precision of the (KoralW + YFSWW) ``Toolbox'' 
for the total
cross-sections of the WW-physics at LEP2 to be {\bf 0.5\%} (preliminary).

\section{Under the carpet}

 From previous sections one may get the impression that our
``Four-Fermion MC Toolbox'' is almost finished. Unfortunately this is not
true, especially for the NC processes. I will present
some of the outstanding problems. They
are of a universal type and can be even
several times bigger than the 0.5\% precision tag, see ref.\
\cite{koralw:1998} for more comments.

{$\bullet$ \bf Numerical instabilities}

The ratio $m^2_{e}/s$ is at LEP2 of the order of $10^{-12}$. 
Together with delicate gauge and unitarity cancellations it leads to
numerical instability problems in the matrix element calculations. Let me
give an explicit example with an event generated by KoralW in the 
$e^-\bar\nu_e \nu_e e^+$ channel:
{\footnotesize
\begin{verbatim}
 pdg            p_x                  p_y                 p_z                  E
  11 -.000000341278492 -.000001614768132  92.983165682216736 92.983165683620868
 -12 -.633000329710363  .109862863634447 -11.985324403772751 12.002531455067720
  12  .771573326908979  .358415826559387 -70.475962939695890 70.481097753801436
 -11 -.138572655920124 -.468277075425702 -10.521878338748095 10.533205107509984
                    four-fermion weight = 913570469940928.500
\end{verbatim}
}
The outgoing electron is highly collinear, the corresponding transfer small, 
and the four-fermion weight (i.e. matrix element) is huge. 
Now, let us modify by hand the last two digits of
 the $p_z$ components of four-momenta and rerun the event:
{\footnotesize
\begin{verbatim}
 pdg               p_x               p_y                 p_z                  E
  11 -.000000341278492 -.000001614768132  92.983165682216722 92.983165683620868
 -12 -.633000329710363  .109862863634447 -11.985324403772731 12.002531455067720
  12  .771573326908979  .358415826559387 -70.475962939695876 70.481097753801436
 -11 -.138572655920124 -.468277075425702 -10.521878338748115 10.533205107509984 
                    four-fermion weight = 25094.3831593953582  
\end{verbatim}
}
The four-fermion weight (matrix element) has changed by 11 orders of
magnitude! 
What can we do about this? We can use a quadruple precision. Our
experience shows that it cures the problems. However it requires
a complex quadruple precision that is not available on all
platforms. Also the speed of the calculation is much lower.
For that and other reasons, specified later on, we prefer another 
solution -- impose additional {\em
post-generation} cut-offs on these dangerous corners of
the phase space:

(1) In the case of the CC-type processes it is only a tiny angular
    cone around the beams, of the size of $10^{-6}$ rad. It influences
    the total cross-section below 0.2\% and can be hidden in 
    the technical precision for the time being. 

(2) In the case of NC-type processes the situation is much worse. There
    are two regions (for final states with the $e^+e^-$ pair) 
    that we had to cut out:
    when invariant mass of produced pairs is smaller than $\sqrt{8}$~GeV
    or when the {\em sum} of transverse momenta squared of 
    visible particles is
    smaller than 600~GeV$^2$. Note that this latter cut
    allows an $e^+e^-$ pair to go to the beam pipe.

{$\bullet$ \bf Limitations of ISR}

The ISR-based description of bremsstrahlung breaks down for
final states with at least one electron (positron) 
collinear to the beam. For such processes the $t$-channel photon
exchange is dominant. In other words the bremsstrahlung is governed by 
$\log \vert t \vert/m^2_{e}$ instead of  
$\log s/m^2_{e}$ . As a result of the missing 
$\log \vert t \vert/s$, too much radiation is generated, especially in the
high-$p_T$ region. Depending on the actual cuts this can significantly
affect the cross-section. 
How can it be cured? 
First of all by adding all the radiation (interferences) missing in 
the soft limit. This means that the 
YFS program should be applied to {\em all six} external particles.
Although radiation in the two $s$- and $t$-channels
separately
is already implemented in MC codes (YFS3 and BHLUMI, respectively),
it is nonetheless not easy technically to merge them into one
coherent algorithm in the case of four-fermion final states. 
(In the case of two fermions, this has already been done in BHWIDE MC code
\cite{bhwide:1997} 
by some of us.)

However, even if we have the radiation 
corrected in the soft limit 
we would still be missing part of the hard photonic
corrections. Can these be dangerous? 
Consider the following event in the LAB frame
($e^+e^-d\bar d$ from KoralW) with a hard transverse photon 
(note that this event passes the cuts defined in the previous section):
{\footnotesize
\begin{verbatim}
                     ================ LAB frame ===================
 pdg               p_x                p_y                p_z                 E
 PHO  5.91545655722838  47.82076396799343   2.67297019944074 48.25932927890828   
 PHO  -.00000027637796   -.00000037152652   -.14955853849146   .14955853849218   
 PHO   .01336462801524    .00882331258921    .02409193017994   .02892896864438   
 PHO   .00000005382246    .00000005171260    .00563331053129   .00563331053178   
   1  -.24118785025702  -1.90677419291787  -2.58051680257066  3.21762744428852
  -1  -.01773699286484   -.01532459077309    .98308780906072   .98341806182419
  11  -.11234211241339    .42661722421839  60.88292736560235 60.88452568487788
 -11 -5.55755400715285 -46.33410540129616 -61.83863527375292 77.47097871243272
                      angles of decay products with resp to beams: 
       d quark          d~ quark            e-                 e+
 -.80199753074017  .99971586568334   .99997374838494   -.79821678134001    
\end{verbatim} }
The problem appears when we transform it to the ``effective frame''. 
In order to calculate the matrix element
that is defined just for four-fermions, the momentum carried out by photons
has to be compensated for and some ``effective'' on-shell electron
``beams'' have to be constructed. After the transformation the event becomes:
{\footnotesize
\begin{verbatim}
                     ============ effective CMS frame =============
 pdg               p_x               p_y                 p_z                  E 
   1 -.008731476493231  .021808545200827  -2.679767691308820  2.679889313144880
  -1 -.016063888417215 -.021968761695611   1.057202898562928  1.057600417800303
  11  .000815509213020 -.001379443999798  66.011587576947306 66.011587598375669
 -11  .023979855697426  .001539660494582 -64.389022784201416 64.389027269943767
                      angles of decay products with resp to beams:  
       d quark           d~ quark              e-                e+
 -.99996157863668    .99966881915163    .99999999970535   -.99999993036524
\end{verbatim} }
As we see, we have created a ``monster'' -- a very ``small angle''
configuration out of a quite transversal event. This leads to a physically
unjustified enhancement in the matrix element calculation, i.e.\ to high
weights or even numerically unstable ones.
What is the solution? Surely, one needs the complete ${\cal O}(\alpha)$
or even higher order photonic corrections. 
This is a very difficult task for the future.
For now we use
temporary fix-up by cutting out the high $p_T$ photons with $\sum p_T^2 \geq
300$~GeV$^2$. Unfortunately, it is not a completely effective cut and on an
occasion numerical instabilities survive.

\section{Summary}

In this talk I presented the ``Four-Fermion MC
Toolbox''. It consists of three MC programs: 
(1) KoralW with
all four-fermion processes, YFS-based ISR and anomalous WWV couplings;
(2) YFSWW with CC03 matrix element, ${\cal O}(\alpha)$ electroweak
corrections to W-pair production, (Initial+WW)-state YFS-based
bremsstrahlung and anomalous WWV couplings;
(3) YFSZZ with NC02 matrix
element, anomalous ZZV couplings and YFS-based ISR.
The current precision of the codes for the WW physics at LEP2 is 2\% for
KoralW and 0.5\% (preliminary) for YFSWW for the total cross section. Based on
this we expect the corresponding total precision of the Toolbox to reach
the gold-plated 0.5\% level soon.
The important limitations that need to be resolved in the future concern
the final states with electrons (positrons) at small angles, that is mostly
the NC-type processes. They include: 
(1) numerical instabilities in matrix element 
(under control by applying cuts or by the use of a quadruple precision);
(2) lack of ``$t$-channel'' bremsstrahlung;
(3) lack of hard-photon matrix element. 
We are on the way to resolving these problems, and
we are looking forward to it with excitement.

\section*{Acknowledgements}
We thank the CERN Theory and EP Divisions and all four LEP Collaborations 
for their support.
This work was supported in part by 
Polish Government grants 
KBN 2P03B08414, 
KBN 2P03B14715, 
the US DoE contract DE-FG05-91ER40627 and DE-AC03-76SF00515,
Maria Sk\l{}odowska-Curie Joint Fund II PAA/DOE-97-316,
and Polish-French Collaboration within IN2P3 through LAPP Annecy.

\section*{References}

\begin{thebibliography}{10}

\bibitem{yr:lep2}
G. Altarelli, T. Sj\"ostrand and F. Zwirner (eds.),
{\em Physics at LEP2} (CERN 96-01, Geneva, 1996), vols.~1 and 2.

\bibitem{yfs:1961}
D.~R. Yennie, S. Frautschi and H. Suura, Ann. Phys. (NY) {\bf 13},  379
  (1961).

\bibitem{jadach:1988}
S. Jadach and B.~F.~L. Ward, Phys. Rev. {\bf D38},  2897  (1988).

\bibitem{jadach:1989}
S. Jadach and B.~F.~L. Ward, Phys. Rev. {\bf D40},  3582  (1989).

\bibitem{yfs2:1990}
S. Jadach and B.~F.~L. Ward, Comput. Phys. Commun. {\bf 56},  351  (1990).

\bibitem{koralw:1995a}
M. Skrzypek, S. Jadach, W. P\l{}aczek and Z. W\c{a}s,
Comput. Phys. Commun. {\bf  94},  216  (1996).

\bibitem{koralw:1995b}
M. Skrzypek {\it et~al.}, Phys. Lett. {\bf B372},  289  (1996).

\bibitem{koralw:1998}
S. Jadach {\it et~al.}, preprint CERN-TH/98-242 (unpublished).

\bibitem{JadachZalewski:1997}
S. Jadach and K. Zalewski, Acta Phys. Polon. {\bf B28},  1363  (1997).

\bibitem{tauola:1992}
M. Je\.zabek, Z. W\c{a}s, S. Jadach and J. H. K\"uhn, Comput. Phys. Commun. {\bf
  70},  69  (1992).

\bibitem{tauola:1993}
R. Decker, S. Jadach, J.~H. K\"uhn and Z. W\c{a}s, Comput. Phys. Commun. {\bf
  76},  361  (1993).

\bibitem{kleiss:1993}
J. Hilgart, R. Kleiss and F.~Le Diberder, Comput. Phys. Commun. {\bf 75},  191
   (1993).

\bibitem{yfsww2:1996}
S. Jadach, W. P{\l}aczek, M. Skrzypek and B.~F.~L. Ward, Phys. Rev. {\bf D54},
   5434  (1996).

\bibitem{yfsww3:1998}
S. Jadach {\it et~al.}, Phys. Lett. {\bf B417},  326  (1998).

\bibitem{yfsww3:1998b}
S. Jadach {\it et~al.}, preprint UTHEP-98-0502, May 1998 (unpublished).

\bibitem{fleischer:1989}
J. Fleischer, F. Jegerlehner and M. Zra\l{}ek, Z. Phys. {\bf C42},  409
  (1989).

\bibitem{kolodziej:1991}
K. Ko\l{}odziej and M. Zra\l{}ek, Phys. Rev. {\bf D43},  3619  (1991).

\bibitem{dittmaier:1992}
S. Dittmaier, M. B\"ohm and A. Denner, Nucl. Phys. {\bf B376},  29  (1992); 
Err.: ibid., {\bf B391}, 483 (1993).

\bibitem{baur:1995}
U. Baur and D. Zeppenfeld, Phys. Rev. Lett. {\bf 75},  1002  (1995).

\bibitem{argyres:1995}
E. Argyres {\it et~al.}, Phys. Lett. {\bf B358},  339  (1995).

\bibitem{koralw:1997}
T. Ishikawa, Y. Kurichara, M. Skrzypek and Z. W\c{a}s, Eur. Phys. J. {\bf C4},
   75  (1998), preprint CERN-TH/97-11.

\bibitem{yfszz:1997}
S. Jadach, W. P{\l}aczek and B.~F.~L. Ward, Phys. Rev. {\bf D56},  6939
  (1997).

\bibitem{fadin:1993}
V.~S. Fadin, V.~A. Khoze and A.~D. Martin, Phys. Lett. {\bf B311},  311
  (1993).

\bibitem{fadin:1994}
V.~S. Fadin, V.~A. Khoze and A.~D. Martin, Phys. Lett. {\bf B320},  141
  (1994).

\bibitem{fadin:1994b}
V.~S. Fadin, V.~A. Khoze and A.~D. Martin, Phys. Rev. {\bf D49},  2247
  (1994).

\bibitem{melnikov:1996}
K. Melnikov and O. Yakovlev, Nucl. Phys. {\bf B471},  90  (1996).

\bibitem{beenakker:1997}
W. Beenakker, A. Chapovskii and F. Berends, Nucl. Phys. {\bf B508},  17
  (1997).

\bibitem{beenakker:1997b}
W. Beenakker, A. Chapovskii and F. Berends, Phys. Lett. {\bf B411},  203
  (1997).

\bibitem{dittmaier:1998}
A. Denner, S. Dittmaier and M. Roth, Nucl. Phys. {\bf B519},  39  (1998).

\bibitem{dittmaier:1998b}
A. Denner, S. Dittmaier and M. Roth, Phys. Lett. {\bf B429},  145  (1998).

\bibitem{bhwide:1997}
S. Jadach, W. P\l{}aczek and B. F. L. Ward, Phys. Lett. {\bf B390}, 298 (1997).

\end{thebibliography}

\end{document}